\documentclass[amsmath,amssymb,pra,twocolumn]{revtex4}
\usepackage[cp1251]{inputenc}
\usepackage[english]{babel}
\usepackage{color}
\usepackage{graphicx}
\usepackage{dcolumn}
\usepackage{braket}
\usepackage{amsmath}

\begin{document}
\bibliographystyle{unsrt}

\title{SPDC once again on the parameters of  transverse entanglement outside the near zone.}
\author{M.V. Fedorov$^{1,2}\,$}
\email{fedorovmv@gmail.com}
\author{S.S. Mernova$^{1,2}$, K.V. Sliporod$^{1,2}$}
\address{$^1$A.M.~Prokhorov General Physics Institute,
 Russian Academy of Sciences, Moscow, Russia\\
 $^2$National Research University Higher School of Economics, Moscow, Russia}

\date{\today}

\begin{abstract}
The reduced density matrix $\rho_r(x,x^\prime)$ of the transverse biphoton state beyond the near zone is found to be real on and only on diagonals in the plane of photon's transverse coordinates $(x,x^\prime)$. This remarkable feature is found to occur at any distances of photon propagation with diffraction spreading completely taken into account. The functions of the reduced density matrix at two orthogonal diagonals, $\rho_r(x,x)$ and $\rho_r(x,-x)$, are considered as the only two complementary single-particle distributions generated by the reduced density matrix. The ratio of widths of these two distributions is interpreted as the new entanglement parameter  $R_{\rm diag}$ which is found to be very close to and completely compatible with the Schmidt entanglement parameter $K$.
\end{abstract}


\maketitle

\section{Introduction}
One of the simplest regimes of Spontaneous Parametric Down-Conversion (SPDC) is the collinear, frequency-degenerate regime with the type-I phase-matching. In this regime the pump propagates in the nonlinear crystal as an extraordinary wave whereas emitted photons are of the ordinary-wave type, and a scheme of the SPDC process is $e\rightarrow o+o$.  The states of SPDC-generated biphotons propagating in a free space  can be entangled in spectral or/and angular variables of emitted photons. Very often these two types of entanglement can be considered independently of each other. In this work we consider (reconsider) only angular or transverse-coordinate entanglement, where ``transverse" means perpendicular to the central propagation direction $Oz$ of both the pump and emitted photons. In fact, this type of entanglement is rather well studied both theoretically and experimentally \cite{Sergienko, Monk, Eberly, Lloid, Vallone, Howell, Kim, Eliel, Anisotr1, Anisotr2, JHE, MVChe}. Parameters characterizing the degree of transverse entanglement are also known rather well, and they are the Schmidt entanglement parameter $K$ \cite{Grobe, CP} and the parameter $R$, defined as the ratio of widths of the unconditional and conditional single-particle photon distributions \cite{R,JPB}. In the near zone close to the exit surface of the crystal where SPDC takes place, both parameters $K$  and $R$ are valid and, at least approximately,  $K\approx R$, both in the momentum (wave-vector) and coordinate representations. It's known also that with the diffraction spreading taken into account the Schmidt parameter $K$ remains constant, i.e., independent of the propagation distance both in the momentum and coordinate representations, and the same is true for the parameter $R$ in the momentum representation. But as shown in the works \cite{JHE} and \cite{MVChe}, beyond the near zone the parameter $R$ found in the coordinate representation becomes dependent on the propagation distance and fails to characterize the degree of transverse entanglement appropriately. This fact is interpreted in \cite{JHE,MVChe} as migration of entanglement to the phase of the biphoton wave function. These results are  reproduced briefly below in Section 3 in the frame of a  double-Gaussian model for the biphoton wave function. But, in our opinion, under the inapplicability  conditions of the traditionally used parameter $R=R_{tr}$ \cite{R,JPB}, the inevitably arising question is whether it's possible to find something instead, i.e. whether there is any other parameter, different from the Schmidt parameter $K$ and having the previous sense of the ratio of widths of some single-particle distributions. In principle, we claim that such a possibility exists, and such single-particle distributions are given by the reduced density matrix on the main- and side-diagonals in the plane of two photon's transverse coordinates. Justification of this assumption is in the following sections of the paper.

\section{General properties of the transverse biphpoton wave functions and density matrices}

Thus, let us consider biphoton states arising in the case of the collinear and frequency degenerate SPDC process with the type-I phase-matching,  with $0z$ being the central propagation axis and with selection by a slit of photons with the wave vectors belonging to the $zx$, perpendicular to the plane containing the crystal optical axis. The wave function of such state in the momentum representation and in near zone is well known to be given by \cite{Monk,Eberly}
\begin{gather}
 \nonumber
 \psi(k_{1x}, k_{2x}) \equiv\psi_{G-s}(k_{1x}, k_{2x})=\\
 \exp\left[-\frac{(k_{1x}+k_{2x})^2 w_p^2}{4}\right]
 \label{wf-k}
 {\rm sinc}\left[\frac{L\lambda_p}{8\pi n_o}(k_{1x}-k_{2x} )^2\right],
\end{gather}
where ${\rm sinc}(u)=\sin u/u$ and $k_{1,2\,x}$ are transverse components ($\perp Oz$) of the photon's wave vectors. The exponential factor on right-hand side of this equation is the transverse profile of the pump with $w_p$ and $\lambda_p$ being its waist and wavelength. The sinc-function characterizes formation of the biphoton beam in the crystal of the length $L$ located in the interval $z\in[-L/2,L/2]$ which excludes appearance in this expression of any additional phase factors. The normalization factors here and henceforth are dropped; $n_o$ is the ordinary-wave refractive index in the crystal.

The wave function (\ref{wf-k}) has the following evident and useful properties: it is real, it does not change if both $k_{1x}$ and $k_{2x}$ change their signs,  and it does not change if $k_{1x}$ and $k_{2x}$ substitute each other, $k_{1x}\rightleftarrows k_{2x}$
\begin{gather}
 \nonumber
  \psi(k_{1x}, k_{2x})=\psi(-k_{1x}, -k_{2x})=\psi(k_{2x}, k_{1x})=\\
 \label{identities}
 =\psi^*(k_{1x}, k_{2x}).
\end{gather}

Beyond the near zone an at longer times $t$, the wave function (\ref{wf-k}) acquires factors, characterizing propagation of photons in a free space along the $z$-axis. :
\begin{gather}
 \nonumber
  \Psi(k_{1x}, k_{2x};\zeta)=\psi(k_{1x}, k_{2x})\times\\
  \nonumber
  \exp\left\{-i\left[ (\omega_1+\omega_2)t-(k_{1z}+k_{2z})z\right]\right\}=\\
 \psi(k_{1x}, k_{2x})\exp\left\{-i\left[\omega_p(t-z/c)+\zeta(k_{1x}^2+k_{2x}^2)\right]\right\},
 \label{wf-k-z}
 \end{gather}
where  $\omega_p=\omega_1+\omega_2$ is the pump frequency, $\omega_1=\omega_2=\omega_p/2$ are frequencies of SPDC photons,  $k_{1,2\,z}\approx \omega_p/2c -\zeta k_{1,2\,x}^2$ in the paraxial approximation, and
\begin{equation}
 \label{zeta}
 \zeta = \frac{z\lambda_p}{2\pi}.
\end{equation}
The wave function (\ref{wf-k-z}) can be used directly for defining the so-called conditional and unconditional single-particle distributions of the probability densities
\begin{equation}
\label{cond-k}
\frac{dW^{\rm(c.)}(k_{1x})}{dk_{1x}}=|\Psi(k_{1x},0;\zeta)|^2\equiv |\psi(k_{1x},0)|^2
\end{equation}
and
\begin{gather}
\nonumber
\frac{dW^{\rm(u.c.)}(k_{1x})}{dk_{1x}}=\int dk_{2x}|\Psi(k_{1x},k_{2x};\zeta)|^2\equiv \\
\label{uncond-k}
\int dk_{2x}|\psi(k_{1x},k_{2x})|^2
\end{gather}
Both of these two distributions are seen to be independent of $\zeta$. For making them comparable with each other, they  should be normalized by the condition that their heights in maxima are equal unity. Then the widths of distributions can be defined as the widths of two curves at the level 0.5, and the ratio of these widths is the traditional ``widths ratio" entanglement parameter $R_{\rm tr}$ of the work  \cite{R}
\begin{equation}
 \label{R-tr-k}
 R_{\rm tr}=\frac{\Delta k^{(u.c.)}_{1x}}{\Delta k^{(c.)}_{1x}}.
\end{equation}

The full and reduced density matrices are defined as
\begin{gather}
 \nonumber
 \rho(k_{1x},k_{2x},k_{1x}^\prime, k_{2x}^\prime;\zeta)=
 \Psi(k_{1x},k_{2x};\zeta)\Psi^*(k_{1x}^\prime,k_{2x}^\prime;\zeta)=\\
 \label{rho-full-k}
 \psi(k_{1x},k_{2x})\psi(k_{1x}^\prime,k_{2x}^\prime)\exp\{i\zeta(k_{1x}^{\prime 2}+k_{2x}^{\prime 2}-k_{1x}^2-k_{2x}^2)\}
\end{gather}
and
\begin{gather}
 \nonumber
 \rho_r(k_{1x},k_{1x}^\prime;\zeta)=\int dk_{2x}\rho(k_{1x},k_{2x},k_{1x}^\prime, k_{2x};\zeta)=\\
  \label{rho-r-k}
  =\int dk_{2x} \psi(k_{1x},k_{2x})\psi(k_{1x}^\prime,k_{2x})
  e^{i\zeta\left(k_{1x}^{\prime\,2}-k_{1x}^2\right)}.
\end{gather}
As follows from this equation, at $\zeta=0$ the reduced density matrix is symmetric:
\begin{equation}
 \label{siim-rho}
 \rho_r(k_{1x},k_{1x}^\prime;\zeta=0)=\rho_r(k_{1x}^\prime,k_{1x};\zeta=0)\equiv \rho_r(k_{1x},k_{1x}^\prime),
\end{equation}
where
\begin{equation}
\label{rho-k-z0}
\rho_r(k_{1x},k_{1x}^\prime)=
\int dk_{2x}\psi(k_{1x},k_{2x})\psi(k_{1x}^\prime,k_{2x}).
\end{equation}

\begin{figure}[h]
  \centering
  \includegraphics[width=5cm]{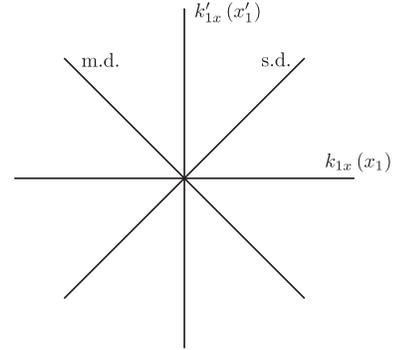}
  \caption{The main ($\rm{m.d.}$) and side ($\rm{s.d.}$) diagonals in the momentum $(k_{1x},k_{1x}^\prime)$- and coordinate $(x_1,x_1^\prime)$-planes. }
  \label{Fig1}
\end{figure}

An even more important consequence of Eq. (\ref{rho-r-k}) is that the reduced density matrix $\rho_r(k_{1x},k_{1x}^\prime;\zeta)$ is real and does not depend on the propagation length $\zeta$ on and only on the diagonals in the $(k_{1x},k_{1x}^\prime)$ plane (see Fig. \ref{Fig1}), i.e. at $k_{1x}=k_{1x}^\prime$ or at $k_{1x}=-k_{1x}^\prime$:
\begin{gather}
 \nonumber
 \left.\rho_r(k_{1x},k_{1x}^\prime;\zeta)\right|_{\rm diag}=\left.\rho_r^*(k_{1x},k_{1x}^\prime;\zeta)\right|_{\rm diag}=\\
 \label{rho-r-k-diag}
 =\left.\rho_r(k_{1x},k_{1x}^\prime)\right|_{\rm diag}.
\end{gather}

Note that this conclusion

 Owing to this, we can introduce the second pair of single-particle distributions represented by the reduced density matrix on the main and side diagonals:
\begin{gather}
 \nonumber
 \frac{dW^{(\rm s.d.)}(k_{1x})}{dk_{1x}}=\rho_r( k_{1x},k_{1x})=\\
 \label{s.d. -k}
 =\int dk_{2x}|\psi(k_{1x},k_{2x})|^2=\frac{dW^{(\rm u.c.)}(k_{1x})}{dk_{1x}}
\end{gather}
 and
\begin{gather}
 \nonumber
 \frac{dW^{\rm (m.d.)}(k_{1x})}{dk_{1x}}=\rho_r( k_{1x},-k_{1x})=\\
 \label{m.d. -k}
 =\int dk_2\psi(k_{1x},k_{2x})\psi(-k_{1x},k_{2x}).
\end{gather}
Both of these distributions do not depend on the photons' propagation length $\zeta$. The side-diagonal distribution is seen to be identical to the unconditional distribution of Eq. (\ref{uncond-k}), whereas the main-diagonal distribution can differ from the conditional one (\ref{cond-k}). If $\Delta k_{1x}^{\rm(s.d.)}$ and $\Delta k_{1x}^{\rm(m.d.)}$ are the widths of the side- and main-diagonal distributions (\ref{s.d. -k}) and (\ref{m.d. -k}), their ratio is the diagonal width-ratio entanglement parameter,
\begin{equation}
\label{R-diag-gen}
 R_{\rm diag}=\frac{\Delta k_{1x}^{\rm(s.d.)}}{\Delta k_{1x}^{\rm(m.d.)}},
\end{equation}
alternative to $R_{\rm tr}$ (\ref{R-tr-k}).

At last, the Schmidt entanglement  parameter $K$ is defined by equations
\begin{gather}
 \nonumber
 \frac{1}{K}=Tr\left(\rho_r^2\right)=\int dk_{1x}dk_{2x}\left[\rho_r(k_{1x},k_{2x})\right]^2=\\
 \nonumber
 \int dk_{1x}dk_{2x}dk_{1x}^\prime dk_{x}^\prime \Psi(k_{1x},k_{2x};\zeta)\Psi^*(k_{1x}^\prime,k_{2x};\zeta)\times\\
 \label{Schm}
 \Psi^*(k_{1x},k_{2x}^\prime;\zeta) \Psi(k_{1x}^\prime,k_{2x}^\prime;\zeta).
\end{gather}
As easily checked, exponential factors in the product of four $\Psi$-functions cancel each other and, hence, $\Psi$-functions (\ref{wf-k-z}) in Eq. (\ref{Schm}) can be replaced everywhere by the Gauss-sinc near-zone wave function $\psi$ (\ref{wf-k}). This means that the Schmidt parameter $K$ remains the same at any values of the propagation distance $\zeta$. Moreover, as we'll see, the Schmidt parameter $K$ is invariant with respect to the change of the  momentum- to coordinate- representation. For this reasons the Schmidt entanglement parameter $K$ can be considered as the benchmark for evaluation of qualities of the parameters $R_{\rm tr}$ (\ref{R-tr-k}) and $R_{\rm diag}$ (\ref{R-diag-gen}).

For transition to the coordinate representation, the wave function (\ref{wf-k-z}) must be Fourier transformed to give
\begin{gather}
 \nonumber
 {\widetilde\Psi}(x_1, x_2;\zeta)=\int dk_{1x}dk_{2x}\Psi(k_{1x},k_{2x};\zeta)e^{i(x_1k_{1x}+x_2k_{2x})}=\\
 \int dk_{1x}dk_{2x}\psi(k_{1x}, k_{2x}) e^{i\zeta(k_{1x}^2+k_{2x}^2)}e^{i(x_1k_{1x}+x_2k_{2x})}.
 \label{wf-x-z}
 \end{gather}
The tilde-symbol $\,\widetilde{}\,$  atop $\Psi$ indicates here and below the wave function in the coordinate representation in order  to differentiate it from $\Psi$ in the momentum presentation, and the same differentiation will be used below for the density matrices  ${\widetilde \rho}$ and $\rho$, for entanglement parameters $\widetilde{R}$ and $R$, etc.

The coordinate wave function ${\widetilde\Psi}(x_1, x_2;\zeta)$ has properties following from the transformation properties (\ref{identities}) of the near-zone momentum wave function (\ref{wf-k}) $\psi(k_{1x},k_{2x})$ (\ref{identities}):
\begin{gather}
 \nonumber
 {\widetilde\Psi}(x_1, x_2;\zeta)={\widetilde\Psi}(-x_1,-x_2;\zeta)={\widetilde\Psi}(x_2, x_1;\zeta)=\\
 \label{properties-x}
 ={\widetilde\Psi}^*(x_1, x_2;-\zeta).
 \end{gather}

The $\zeta$-dependent conditional and unconditional probability densities in the coordinate representation determined by the wave function of Eq. (\ref{wf-x-z}) are given by
\begin{gather}
\nonumber
\frac{d\widetilde{W}^{\rm{(c.)}}(x_1)}{dx_{1}}=\left|{\widetilde\Psi}(x_1, x_2=0;\zeta)\right|^2=\\
\nonumber
 \int dk_{1x}dk_{1x}^\prime dk_{2x} dk_{2x}^\prime\psi(k_{1x},k_{2x})\psi(k_{1x}^\prime,k_{2x}^\prime)e^{i(k_{1x}-k^\prime_{1x})x_1}\times\\
 e^{i\zeta (k_{1x}^2+k_{2x}^2-k_{1x}^{\prime\,2}-k_{2x}^{\prime\,2})}
\label{cond-x-}
\end{gather}
and
\begin{gather}
\nonumber
\frac{d\widetilde{W}^{\rm{(u.c.)}}(x_1)}{dx_1}\equiv\frac{d\widetilde{W}^{\rm{(s.d.)}}(x_1)}{dx_1}=\int dx_2\left|\Psi(x_1,x_2;\zeta)\right|^2=\\
\nonumber
= \int dk_{1x}dk_{1x}^\prime dk_{2x} \psi(k_{1x},k_{2x})\psi(k_{1x}^\prime,k_{2x})\times\\
 \nonumber
 e^{i\zeta (k_{1x}^2-k_{1x}^{\prime\,2})}e^{i(k_{1x}-k^\prime_{1x})x_1}=\\
 =\int dk_{1x}dk_{1x}^\prime \rho_r(k_{1x},k_{1x}^\prime;\zeta)e^{i(k_{1x}-k^\prime_{1x})x_1}.
\label{uncond-x-}
\end{gather}
The full and reduced density matrices in the coordinate representation are defined as
\begin{gather}
 \nonumber
 \widetilde{\rho}(x_1,x_2,x_1^\prime,x_2^\prime;\zeta)=
  \widetilde{\Psi}(x_1,x_2;\zeta)\widetilde{\Psi}^*(x_1^\prime,x_2^\prime;\zeta)=\\
  \nonumber
  \int dk_{1x}dk_{2x}dk_{1x}^\prime dk_{2x}^\prime e^{i[x_1k_{1x}+x_2k_{2x}-x_1^\prime k_{1x}^\prime
 -x_2^\prime k_{2x}^\prime]}\\
  e^{i\zeta[-k_{1x}^2-k_{2x}^2+k_{1x}^{\prime\,2}+k_{2x}^{\prime\,2}]}
  \psi(k_{1x}, k_{2x})\psi^*(k_{1x}^\prime, k_{2x}^\prime)
  \label{rho-x}
\end{gather}
and
\begin{gather}
 \nonumber
 \widetilde{\rho}_r(x_1,x_1^\prime;\zeta)=\int dx_2 \widetilde{\rho}(x_1,x_2,x_1^\prime,x_2;\zeta)=\\
   \label{near-far rel}
   \int dk_{1x}dk_{1x}^\prime e^{i(x_1k_{1x}-x_1^\prime k_{1x}^\prime)}
  e^{i\zeta(-k_{1x}^2+k_{1x}^{\prime\,2})}
  \rho_r(k_{1x},k_{1x}^\prime).
\end{gather}

As follows from the properties (\ref{identities}) of the wave function $\psi(k_{1x},k_{1x}^\prime)$ (\ref{wf-k}) and from the definition (\ref {rho-k-z0}) of $\rho_r(k_{1x},k_{1x}^\prime)$, the properties of $\rho_r(k_{1x},k_{1x}^\prime)$ are very similar to those of the wave function $\psi (k_{1x},k_{ 1x}^\prime)$: the function $\rho_r(k_{1x},k_{1x}^\prime)$ is real, it does not change when both of its arguments  $k_{1x} $ and $k_{1x} ^\prime$ change signs, and also it does not change after permutation of its arguments $k_{1x}\rightleftarrows k_{1x}^\prime$. With these notes taken into account, we find the $\zeta$-dependent  coordinate reduced density matrix ${\widetilde\rho}(x_1,x_1^\prime;\zeta)$ (\ref{near-far rel})  on diagonals in the $(x_1,x_1^\prime)$ plane (Fig. 1), which are given by
\begin{gather}
 \nonumber
 \widetilde{\rho}_r^{\,\rm(s.d.)}=\widetilde{\rho}_r(x_1,x_1;\zeta)=\\
  \int dk_{1x}dk_{1x}^\prime e^{ix_1(k_{1x}-k_{1x}^\prime)}
   e^{i\zeta [-k_{1x}^2+k_{1x}^{\prime\,2}]}
   \rho_r(k_{1x},k_{1x}^\prime)
 \label{rho-r-s.d.}
\end{gather}
and
 \begin{gather}
 \nonumber
 \widetilde{\rho}_r^{\,\rm(m.d.)}=\widetilde{\rho}_r(x_1,-x_1;\zeta)=\\
  \int dk_{1x}dk_{1x}^\prime e^{ix_1(k_{1x}+k_{1x}^\prime)}
   e^{i\zeta[-k_{1x}^2+k_{1x}^{\prime\,2}]}
   \rho_r(k_{1x},k_{1x}^\prime).
 \label{rho-r-m.d.}
\end{gather}

It's easy to find such manipulations with the integration variables  in   Eqs. (\ref{rho-r-s.d.}) and (\ref{rho-r-m.d.}) which convert these expressions to themselves but with complex conjugation, and this  means that at diagonals expressions  (\ref{rho-r-s.d.}) and (\ref{rho-r-m.d.}) are real.  For the side diagonal, this manipulation is simply the permutation of $k_{1x}$ and $k_{1x}^\prime$ which gives
\begin{equation}
 \widetilde{\rho}_r(x_1,x_1;\zeta)\,^{^{k_{1x}\rightleftarrows k_{1x}^\prime}_{\quad_{==}}}\,\widetilde{\rho}_r^{\,*}(x_1,x_1;\zeta)
\end{equation}
In the case of the main  diagonal the permutation of $k_{1x}$ and $k_{1x}^\prime$ must be supplemented by changing signs of these variables ($k_{1x}\rightarrow -k_{1x}$ and $k_{1x}^\prime\rightarrow -k_{1x}^\prime $, which gives the same result
\begin{equation}
 \widetilde{\rho}_r(x_1,-x_1;\zeta)\,^{^{k_{1x}\rightleftarrows -k_{1x}^\prime}_{\quad_{  ==}}}\,\widetilde{\rho}_r^{\,*}(x_1,-x_1;\zeta).
\end{equation}
Thus, once again, at both diagonals the $\zeta$-dependent coordinate reduced density matrices $\widetilde{\rho}_r(x_1,x_1^\prime;\zeta)$ are real. Moreover, in a general case $\zeta\neq 0$,  ${\rm Im}[{\widetilde \rho_r(x_1,x_1^\prime;\zeta)}]= 0$ only at these two directions, $\rm m.d.$ and $\rm s.d.$, and otherwise ${\rm Im}[{\widetilde \rho_r(x_1,x_1^\prime;\zeta)}]\neq 0$. Note that though this result follows directly from the properties (\ref{identities}) of the near-zone sinc-Gaussian wave function $\psi_{G-s}(k_{1x},k_{2x})$ (\ref{wf-k}), there is a much wider class of the near-zone wave functions $\psi(k_{1x}+k_{2x},k_{1x}-k_{2x})$, differing from $\psi_{G-s}$ but obeying the same properties (\ref{identities}) and, hence, generating reduced density matrices with zero imaginary parts on diagonals. But for the goals of the present work, consideration of the near-zone sinc-Gaussian wave function $\psi_{G-s}(k_{1x},k_{2x})$ (\ref{wf-k}) is sufficient.

Importance of zeroing imaginary parts of the reduced density matrix at diagonals is related to a possibility of considering the real  functions of a single argument $x_1$, $\widetilde{\rho}_r(x_1,x_1;\zeta)$ and $\widetilde{\rho}_r(x_1,-x_1;\zeta)$, as the only well defined single-particle distributions generated by the reduced density matrix. The widths of these distributions, $\Delta x_1^{\,(\rm s.d.)}$ and $\Delta x_1^{\,(\rm m.d.)}$, can be used for defining the fully ``diagonal$"$ width-ratio parameter characterizing the degree of entanglement in the coordinate representation
\begin{equation}
 \label{R diag}
  {\widetilde R}_{\rm diag}=\frac{\Delta x_1^{\,(\rm s.d),}}{\Delta x_1^{\,(\rm m.d.)}}.
\end{equation}

Note that in accordance with Eqs. (\ref{rho-x}), (\ref{near-far rel}), in terms of the $\zeta$-dependent coordinate wave function $\widetilde{\Psi}(x_1,x_2;\zeta)$ (\ref{wf-x-z}), the $\rm s.d.$- and $\rm m.d.$-reduced density matrices can be written as
\begin{equation}
 \widetilde{\rho}_r(x_1,x_1;\zeta)=\widetilde{\rho}_r^{\,(\rm s.d.)}=\int dx_2|\widetilde{\Psi}(x_1,x_2;\zeta)|^2
 \label{single-sd}
\end{equation}
and
\begin{gather}
 \nonumber
 \widetilde{\rho}_r(x_1,-x_1;\zeta)=\widetilde{\rho}_r^{\,(\rm m.d.)}=\\
 =\int dx_2\widetilde{\Psi}(x_1,x_2;\zeta)
 \widetilde{\Psi}^{*}(-x_1,x_2;\zeta)
 \label{single-md}
\end{gather}
The first of these two expressions, (\ref{single-sd}), shows that the distribution $\widetilde{\rho}_r^{\,(\rm s.d.)}(x_1,x_1)$ coincides with the distribution which is considered usually as the unconditional one in terms of the wave function $\widetilde{\Psi}(x_1,x_2;\zeta)$, and hence $\Delta x_1^{\,(\rm s.d.)}\equiv \Delta x_{1\,\Psi}^{(\rm u.c.)}$.

As for the distribution $\widetilde{\rho}_r^{\,(\rm m.d.)}(x_1,-x_1;\zeta)$, it does not have such simple interpretation. But it can be considered as a substitute of the usual conditional distribution $|\widetilde{\Psi}(x_1,x_2=0;\zeta)|^2$ when the latter gives evidently wrong results. As explained above in the Introduction and as shown in the works \cite{Eberly,MVChe}, at $zeta\neq 0$ the conditional single-particle distribution $|\widetilde{\Psi}(x_1,x_2=0;\zeta)|^2$ and its width cannot be used for correct definition of the entanglement parameter ${\widetilde R}$ because  ``entanglement migrates to the phase of the wave function$"$. In contrast to this, the distribution $\widetilde{\rho}_r^{\,(\rm m.d.)}(x_1,-x_1;\zeta)$ is phase-free, because of which it cannot demonstrate any migration of entanglement, and in this sense it is preferable.

As for the Schmidt parameter, as mentioned above, it is invariant with respect to the transition from the momentum to coordinate representations, and it is independent of the photon's propagation length $\zeta$, i.e. not affected by diffraction. For these reasons the parameter $K$ remains the main benchmark for evaluation of quality of other parameters such as $R_{\rm tr}$, ${\widetilde R}_{\rm tr}$, $R_{\rm diag}$ and ${\widetilde R}_{\rm diag}$.

In the following two sections features of entanglement parameters are analyzed in more details for the cases of the model double-Gaussian wave function analytically (section 3) and numerically for the Gauss-sinc wave function (\ref{wf-k}) (section 4).

\section{The model double-Gaussian wave function}

\begin{figure}[h]
  \centering
  \includegraphics[width=7cm]{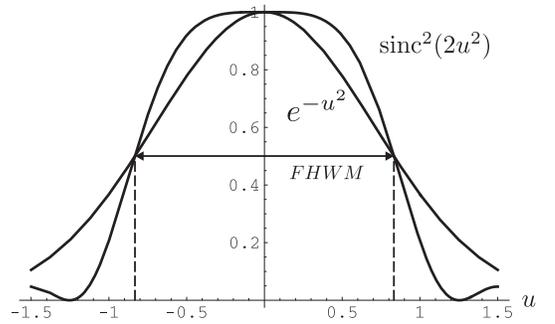}
  \caption{The functions $\exp(-u^2)$ and ${\rm sinc^2}(2u^2)$ and coincidence of their widths at half-maxima.}
  \label{Fig2}
\end{figure}

The near-zone model double-Gaussian wave function in the momentum representation arises when the sinc-function in Eq. (\ref{wf-k}) is substituted by the Gaussian function to give
\begin{gather}
 \nonumber
 \psi_{\rm 2G}(k_{1x},k_{2x})
 \propto\exp\left[-\frac{a^2(k_{1x}+k_{2x})^2}{2}\right]\times\\
 \label{wf-mom-zeta=0}
  \exp\left[-\frac{b^2(k_{1x}-k_{2x})^2}{2}\right].
\end{gather}
where $a=w/\sqrt{2}$ and $b=\sqrt{L\lambda_p}/(4{\sqrt{\pi n_o}})$. With this definition of the parameter $b$, the sinc-function in Eq. (\ref{wf-k}) takes the form ${\rm sinc}[2b^2(k_{1x}-k_{2x})^2]$ and the coefficient 2 provides equality of  the Full Widths at Half-Maxima of the sinc$^2$- and the squared second Gaussian function in Eq. (\ref{wf-mom-zeta=0}): ${\rm FWHM}\left[{\rm sinc^2}(2u^2)\right]\approx{\rm FHWM}\left[\exp(-u^2)\right]$, where $u=b(k_{1x}-k_{2x})$ (see Fig.{\ref{Fig2}}) .

Below we will use widely the definitions of the characteristic length $\delta x=\sqrt{ab}$, dimensionless transverse wave vectors $q_{1,2}= \delta x\times k_{1,2\,x}$ and the dimensionless control parameter
\begin{equation}
 \label{eta}
 \eta=\frac{b}{a},
\end{equation}
in terms of which $a=\delta x/\sqrt{\eta}$, $b=\delta x\sqrt{\eta}$ and Eq.(\ref{wf-mom-zeta=0}) takes the simplest form
\begin{gather}
 \psi_{\rm 2G}(q_1,q_2)
 \propto\exp\left[-\frac{(q_1+q_2)^2}{2\eta}
 \label{wf-mom-eta-Z=0}
  -\eta\frac{(q_1-q_2)^2}{2}\right].
\end{gather}
The double-Gaussian wave function of Eqs. (\ref{wf-mom-zeta=0}) and (\ref{wf-mom-eta-Z=0}) is known to yield the Schmidt entanglement parameter (\ref{Schm}) of the form
\begin{equation}
 \label{K-2G}
 K_{\rm 2G}=\frac{a^2+b^2}{2ab}=\frac{1+\eta^2}{2\eta}.
\end{equation}
Note that the dimensionless form of the Schmidt entanglement parameter demonstrates most clearly the mentioned above symmetry of the double-Gaussian wave function (\ref{wf-mom-zeta=0}), (\ref{wf-mom-eta-Z=0}) owing to which the Schmidt parameter $K_{\rm 2G}$ appears to be invariant with respect to  the variable substitution $\eta\rightarrow 1/\eta$ and, besides, $K_{\rm 2G}$ is minimal and equals 1 (no entanglement) at $\eta=1$ :
\begin{equation}
 \label{symmetry}
 K_{\rm 2G}(\eta)\equiv K_{\rm 2G}(1/\eta)\;{\rm and}\; K_{\rm 2G\,min}=K_{\rm 2G}(1)=1.
\end{equation}

In addition to the Schmidt parameter $K_{\rm 2G}(\eta)$ (\ref{symmetry}), we easily find also the unconditional and conditional single-particle distributions determined by the wave function $\psi_{\rm 2G}(k_{1x},k_{2x}$ (\ref{wf-mom-zeta=0}), (\ref{wf-mom-eta-Z=0})
\begin{gather}
 \nonumber
 \frac{dW_{\rm 2G}^{\rm (u.c.)}}{dk_{1x}}=\int dk_{2x} |\Psi_{\rm 2G}(k_{1x},k_{2x})|^2=\\
 \label{uncond-2G=z=0}
 \exp\left\{-\frac{4a^2b^2}{a^2+b^2}k_{1x}^2\right\}=\exp\left\{-\frac{4\eta^2}{1+\eta^2}\,q_1^2\right\}
\end{gather}
and
\begin{gather}
 \nonumber
 \frac{dW_{\rm 2G}^{\rm (c.)}}{dk_{1x}}
 =|\Psi_{\rm 2G}(k_{1x},0)|^2= \\
 \label{cond-2G-z=0}
 \exp\left\{-(a^2+b^2)k_{1x}^2\right\}=\exp\left\{-(1+\eta^2)\,q_1^2\right\}.
\end{gather}
The widths of these two single-particle distributions are
\begin{equation}
  \label{widths-k-z=0}
 \Delta k_{1x}^{\rm (u.c.)}=\frac{a^2+b^2}{2ab}\quad{\rm and}\quad\Delta k_{1x}^{\rm (c.)}=\frac{1}{\sqrt{a^2+b^2}}
 \end{equation}
 The ratio of these widths is the traditionally used width-ratio parameter of Ref. \cite{R}, in the momentum representation coinciding with the Schmidt entanglement parameter:
\begin{equation}
 \label{R-tr-near-k}
 R_{\rm tr\,{\rm 2G}}= \frac{\Delta k_{1x}^{\rm (u.c.)}}{\Delta k_{1x}^{\rm (c.)}}=\frac{a^2+b^2}{2ab}=\frac{1+\eta^2}{2\eta}=K_{\rm 2G}.
\end{equation}

The double-Gaussian, near-zone reduced density matrix $\rho_{r\,{\rm 2G}}$ is also found easily from  its general definition (\ref{rho-k-z0}) and Eqs. (\ref{wf-mom-zeta=0}), (\ref{wf-mom-eta-Z=0})  for the wave function
\begin{gather}
 \nonumber
 \left.\rho_{r\,{\rm 2G}}\right|_{z=0}\propto\\ \exp\left\{-\frac{a^2b^2}{a^2+b^2}(k_{1x}+k_{1x}^\prime)^2-
 \nonumber
 \frac{a^2+b^2}{4}(k_{1x}-k_{1x}^\prime)^2\right\}=\\
 \label{red-near-2G}
 \exp\left\{-\frac{\eta^2}{1+\eta^2}(q_1+q_1^\prime)^2-
 \frac{1+\eta^2}{4}(q_1-q_1^\prime)^2\right\}.
\end{gather}

Each of two terms in these exponential functions determines the single-particle distributions along diagonals of Fig. \ref{Fig1}:
\begin{equation}
\label{s.d.-k-2G}
 \frac{dW_{\rm 2G}^{(\rm s.d.)}}{dk_{1x}}=\rho_{r\,{\rm 2G}}(k_{1x},k_{1x)}=
 \exp\left\{-\frac{4a^2b^2}{a^2+b^2}k_{1x}^2\right\}
\end{equation}
and
\begin{equation}
 \label{m.d.-k-2G}
 \frac{dW_{\rm 2G}^{(\rm m.d.)}}{dk_{1x}}=\rho_{r\,{\rm 2G}}(k_{1x},-k_{1x)}=
 \exp\left\{-(a^2+b^2)k_{1x}^2\right\}.
\end{equation}
Obviously, the diagonal distributions (\ref{s.d.-k-2G}) and (\ref{m.d.-k-2G}) generated by the reduced density matrix (\ref{red-near-2G}) exactly coincide, respectively, with the unconditional (\ref{uncond-2G=z=0}) and conditional (\ref{cond-2G-z=0}) distributions generated by the wave function (\ref{wf-mom-zeta=0}). Consequently, the same is true for the widths of these distributions, $\Delta k_{1x}^{\rm (s.d.)}$ and $\Delta k_{1x}^{\rm (m.d.)}$ which coincide, respectively, with
 $\Delta k_{1x}^{\rm (u.c.)}$ and $\Delta k_{1x}^{\rm (c.)}$. The ratio of widths of the diagonal single-particle distributions is the diagonal entanglement parameter (\ref{R-diag-gen}), which appears to be identical in the case under consideration to two other entanglement parameters(\ref{R-tr-near-k})
\begin{equation}
 \label{2G-ent-params-k}
 R_{\rm diag\,2G}=\frac{\Delta k_{1x}^{\rm (s.d.)}}{\Delta k_{1x}^{\rm (m.d.)}}=R_{\rm tr\,2G}=K_{\rm 2G}=\frac{a^2+b^2}{2ab}.
\end{equation}

Beyond the near zone, the wave functions (\ref{wf-mom-zeta=0}), (\ref{wf-mom-eta-Z=0}) and the reduced density matrix (\ref{red-near-2G}) acquire additional propagation-factors determined by Eqs. (\ref{wf-k-z}) and (\ref{near-far rel}). As the result, the $\zeta$-depended wave function $\Psi_{\rm 2G}(k_{1x}, k_{2x};\zeta)$ and the reduced density matrix $\Psi_{\rm 2G}( k_{1x}, k_{1x},k_{2x};\zeta)$ take the forms
\begin{gather}
 \nonumber
 \Psi_{\rm 2G}(k_{1x},k_{2x};\zeta)=\exp\left[-\frac{(a^2+i\zeta)(k_{1x}+k_{2x})^2}{2}\right.-\\
 \label{wf-2G-mom-z}
  \left.-\frac{(b^2+i\zeta)(k_{1x}-k_{2x})^2}{2}\right]
\end{gather}
and
\begin{equation}
\label{propg-psi}
 \rho_{r\,\rm 2G}(k_{1x},k_{1x}^\prime;\zeta)=\left.\rho_{r\,\rm 2G}(k_{1x},k_{1x}^{\,\prime})\right|_{\zeta=0}
 e^{i\zeta(k_{1x}^{\prime\, 2}-k_{1x}^2)}
\end{equation}
with $\left.\rho_{r\,\rm 2G}(k_{1x},k_{1x}^{\,\prime})\right|_{\zeta=0}$ given by Eq. (\ref{red-near-2G}).

\begin{figure}[h]
  \centering
  \includegraphics[width=7cm]{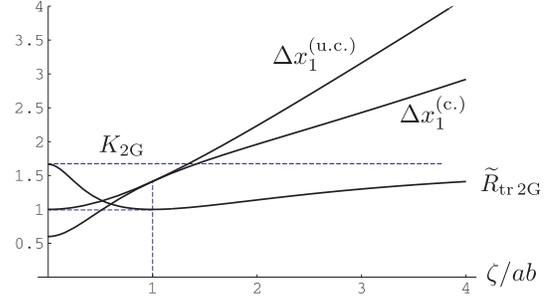}
  \caption{Found with the use of the wave function ${\widetilde \Psi}_{\rm 2G}(x_1,x_2;\zeta)$ (\ref{wf-coord-zeta}) widths of the unconditional and conditional single-particle distributions and their ratio - parameter ${\widetilde R}_{\rm tr\,2G}(\zeta)$ (\ref{R-x-zeta}) vs. $\zeta/ab$; $\Delta x_1^{(u.c.)}$ and $\Delta x_1^{(c)}$ are in units of $\sqrt{a^2+b^2}$ and $a/b=3$; the dashed line is the Schmidt entanglement parameter $K_{\rm 2G}$ (\ref{K-2G}).}\label{Fig3}
\end{figure}

These phase factors do not affect any of the above described single-particle distributions (\ref{uncond-2G=z=0}),  (\ref{cond-2G-z=0}), (\ref{s.d.-k-2G}),(\ref{m.d.-k-2G}) and any of the entanglement parameters (\ref{2G-ent-params-k}) in the momentum representation. But they do affect the same characteristics of the diverging biphoton beams in the coordinate representation. The coordinate wave function is obtained with the help of the double Fourier transformation from
$\Psi_{\rm 2G}(k_{1x},k_{2x};\zeta)$ of Eqs. (\ref{wf-mom-zeta=0}), (\ref{propg-psi}) which gives
\begin{equation}
 \widetilde{ \Psi}_{\rm 2G}(x_1,x_2;\zeta)
 \propto\exp\left[-\frac{(x_1+x_2)^2}{8(a^2+i\zeta)}\right]\\
 \label{wf-coord-zeta}
  \exp\left[-\frac{(x_1-x_2)^2}{8(b^2+i\zeta)}\right].
\end{equation}
The unconditional single-particle probability distribution determined by this wave function can be presented in the form
\begin{gather}
 \nonumber
 \frac{dW^{(\rm u.c.)}(x_1)}{dx_1}=\int dx_2|\widetilde{ \Psi}_{\rm 2G}(x_1,x_2;\zeta)|^2=\\
 \label{2G-uncond-x}
 =\exp\left\{-x_1^2/\left[\Delta x_1^{(\rm u.c.)}(\zeta)\right]^2\right\},
\end{gather}
where the calculated width of the unconditional single-particle distribution is given by
\begin{equation}
 \label{width-u.c.-zeta}
 \Delta x_1^{(\rm u.c.)}(\zeta)= \frac{\sqrt{(a^2+b^2)(a^2b^2+\zeta^2)}}{ab}.
\end{equation}
Another single-particle distribution generated by the wave function (\ref{wf-2G-mom-z}) is the conditional one:
\begin{gather}
 \nonumber
 \frac{dW^{\rm (c.)}(x_1)}{dx_1}= |\Psi_{\rm 2G}(x_1,0;\zeta)|^2=\\
 \label{2G-cond-x}
 \exp\left\{-x_1^2/\left[\Delta x_1^{(\rm c.)}(\zeta)\right]^2\right\},
\end{gather}
with the  width of the conditional single-particle distribution given by
\begin{equation}
 \label{width-cond-zeta}
 \Delta x_1^{(\rm c.)}(\zeta)=
 2\sqrt{\frac{(a^4+\zeta^2)(b^4+\zeta^2)}{(a^2+b^2)(a^2 b^2+\zeta^2)}},
\end{equation}
By definition \cite{R} the traditionally used width-ratio parameter determining is the ratio of unconditional to conditional widths
\begin{equation}
 \label{R-x-zeta}
 {\widetilde R}_{\rm tr\,2G}(\zeta) =\frac{(a^2+b^2)}{2ab}\frac{a^2b^2+\zeta^2}{\sqrt{(a^4+\zeta^2)(b^4+\zeta^2)}}.
\end{equation}
Both unconditional and conditional widths and the parameter ${\widetilde R}_{\rm tr\, 2G}$ are shown in Fig. \ref{Fig3} in their dependence on the normalized propagation length $\zeta$
This picture and equations (\ref{width-u.c.-zeta})-(\ref{R-x-zeta}) agree with the results and conclusions of the works \cite{JHE,MVChe}, and they show that the parameter ${\widetilde R}_{\rm tr\,2G}(\zeta)$ strongly depends on $\zeta$ and significantly deviates from the Schmidt parameter $K_{\rm 2G}={\rm const.}$, owing to which it cannot be considered as the transverse entanglement quantifier if only $\zeta\neq 0$.

An alternative approach to the entanglement quantification is related to the use of the reduced density matrix in the coordinate representation, which is defined as $\rho_r(x_1,x_1^\prime;\zeta)=\int dx_2{\widetilde \Psi}(x_1,x_2;\zeta){\widetilde \Psi}^*(x_1^\prime,x_2;\zeta)$ with the wave function of Eq. (\ref{wf-coord-zeta}), and after a series of integrations it can be reduced to the form
\begin{gather}
 \nonumber
 \rho_r(x_1,x_1^\prime;\zeta)=\exp\left\{-\frac{(x_1+x_1^\prime)^2}{4\left[\Delta x_1^{(\rm s.d.)}\right]^2}\right.\\
 \label{red-x-va-widths}
 -\left.\frac{(x_1-x_1^\prime)^2}{4\left[\Delta x_1^{(\rm m.d.)}\right]^2}-i\zeta
 \frac{x_1^2-x_1^{\prime\,2}}{4(a^2b^2+\zeta^2)}\right\},
\end{gather}
where $\Delta x_1^{(\rm m.d.)}$ and $\Delta x_1^{(\rm s.d.)}$ are width of single-particle probability distributions along the main and side diagonals in the $(x_1,x_1^\prime)$ plane
\begin{equation}
 \label{widths-m}
 \Delta x_1^{(\rm m.d.)}(\zeta)=2\sqrt{\frac{a^2b^2+\zeta^2}{a^2+b^2}}\neq\Delta x_1^{\rm(c.)}(\zeta)
\end{equation}
and
\begin{equation}
 \label{widths-s}
  \Delta x_1^{(\rm s.d.)}(\zeta)=\frac{\sqrt{(a^2+b^2)(a^2b^2+\zeta^2)}}{ab}=\Delta x_1^{\rm(u.c.)}(\zeta).
\end{equation}
Note once again, that if the side-diagonal width $\Delta x_1^{(\rm s.d.)}(\zeta)$ (\ref{widths-s}) coincides with the unconditional width $\Delta x_1^{(\rm u.c.)}(\zeta)$ (\ref{width-u.c.-zeta}) at any values of $\zeta$, i.e. both in near zone and beyond it, the  main-diagonal width $\Delta x_1^{(\rm m.d.)}(\zeta)$ (\ref{widths-m}) coincides with the conditional width $\Delta x_1^{(\rm c.)}(\zeta)$ (\ref{width-cond-zeta})  only at $\zeta=0$, whereas beyond the near zone $\Delta x_1^{(\rm m.d.)}(\zeta)$ differs significantly from  $\Delta x_1^{(\rm u.c.)}(\zeta)$. The ratio of two widths (\ref{widths-s}) and (\ref{widths-m})  gives the diagonal entanglement parameter for the biphoton states in the coordinate representation.

\begin{equation}
 \label{R-diag-2G}
  \widetilde{R}_{\rm diag\,2G}(\zeta)=
  \frac{\Delta x_1^{(\rm s.d.)}}{\Delta x_1^{(\rm m.d.)}}=
 \frac{a^2+b^2}{2ab}=K_{\rm 2G}.
\end{equation}

\begin{figure}[h]
  \centering
  \includegraphics[width=8cm]{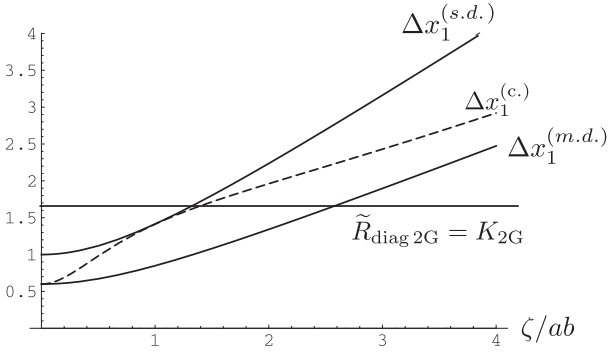}
  \caption{The widths $\Delta x_1^{\rm(s.d.)}$ and $\Delta x_1^{\rm (m.d.)}$ vs. $\zeta$ and their ratio $R_{\rm diag\, 2G}=\Delta x_1^{(\rm s.d.)}/\Delta x_1^{(\rm m.d.)}$; for comparison the dash line shows the usual conditional-distribution width $\Delta x_1^{\rm(c.)}(\zeta)$ (\ref{width-cond-zeta}) which differs significantly from $\Delta x_1^{\rm (m.d.)}(\zeta)$; as well as in Fig.2, the ratio $a/b$ is taken equal 3, and all widths are in units of $\sqrt{a^2+b^2}$.}\label{Fig4}
\end{figure}

 Two diagonal widths $\Delta x_1^{(\rm s.d.)}(\zeta)$ and $\Delta x_1^{(\rm m.d.)}(\zeta)$ are shown in Fig.\ref{Fig4} together with the entanglement parameters $\widetilde{R}_{\rm diag\,2G}$ and $K_{\rm 2G}$.
Remarkably enough, the parameter $\widetilde{R}_{\rm diag\,2G}(\zeta)$ (\ref{R-diag-2G}) remains constant at any values of $\zeta$ and coincides exactly with the Schmidt entanglement parameter $K_{\rm 2G}$ (\ref{K-2G}), in contrast to the traditional width-ratio parameter ${\widetilde R}_{\rm tr\, 2G}(\zeta)$ which appears to be inapplicable for characterization of transverse entanglement beyond the near zone.

\section{Sinc-Gaussian wave function. Numerical simulations.}

The given above analysis based on the use of the model double-Gaussian  wave function (\ref{wf-mom-zeta=0}), (\ref{wf-mom-eta-Z=0}), (\ref{wf-coord-zeta}) shows clearly that in this case the diagonal  parameter of entanglement ${\widetilde R}_{\rm diag\,2G}(\zeta)$ (\ref{R-diag-2G}) is a perfectly good substitute of the traditional width-ratio parameter ${\widetilde R}_{\rm tr\,2G}(\zeta)$ when the latter significantly deviates from $K_{\rm 2G}$ and appears to be inapplicable for characterization of the degree of entanglement. In contrast to this ${\widetilde R}_{\rm diag\,2G}(\zeta)=const.=K_{\rm 2G}$. However the remaining question is how sensitive is this conclusion to modeling the true transverse Gauss-sinc biphoton wave function $\Psi_{\rm G-s}$ (\ref{wf-k}) by the model double-Gaussian wave function $\Psi_{\rm 2G}$ (\ref{wf-mom-zeta=0}). Unfortunately, it's hardly possible to answer this question by means of any analytical derivations and, obviously, numerical solutions are needed. The main results of such solutions are presented below.

But first, we note that, as before, the quality of traditional and diagonal entanglement parameters will be evaluated by comparing them with the Schmidt parameter $K_{\rm Gs}$, and the latter will be considered as a kind of benchmark parameter of the degree of entanglement. To meet this goal, the parameter $K_{\rm Gs}(\eta)$ must have properties close to those of the parameter $K_{\rm 2G}(\eta)$, the main of which is symmetry with respect to the replacement $\eta \rightarrow 1/\eta$ (\ref{symmetry}). But the parameter $K_{\rm G-s}(\eta)$ does not necessarily satisfy this condition automatically. On the other hand, there is some uncertainty in extracting the parameter $b$ from the general expression (\ref {wf-k}) for the Gauss-sinc wave function  $\Psi_{\rm G-s}$. Indeed, we can redefine the sinc function in the general expression (\ref{wf-k}) for $\Psi_{\rm G-s}$ as
\begin{equation}
 \label{s-parameter}
 {\rm sinc}_s(k_{1x}-k_{2x})={\rm sinc}\left(s*b^2(k_{1x}-k_{2x})^2\right),
\end{equation}
where $s$ is the fitting parameter, while
\begin{equation}
 \label{b}
 b=\sqrt{\frac{L\lambda_p}{s*8\pi n_{\rm o}}}
\end{equation}
remains one of two parameters in the definition of the control parameter $\eta=b/a $ (\ref{eta}).
With these redefinitions, the Gauss-sinc wave function (\ref{wf-k}) takes the form
\begin{gather}
 \nonumber
 \psi_s(k_{1x},k_{2x})=\exp\left(-a^2\frac{(k_{1x}+k_{2x})^2}{2}\right){\rm sinc}_s(k_{1x}-k_{2x})=\\
\label{Psi-s}
\exp\left[-\frac{(q_1+q_2)^2}{2\eta}\right]  {\rm sinc}\left[s*\eta(q_1-q_2)^2\right],
\end{gather}
where, as previously, $q_{1,2}=k_{1,2 x}/\delta x$ and $\delta x=\sqrt{ab}$. With the wave function $\psi_s$ (\ref{Psi-s}) used instead of $\Psi$ in Eq. (\ref{Schm}) we have calculated numerically the Schmidt entanglement parameter $K_{\rm G-s}(\eta)$ at various values of $s$ and $\eta$ and found the optimal value of $s$ providing location of the minimum of $K_{\rm G-s}(\eta)$ at $\eta=1$. This optimal value of the fitting parameter appeared to be $s_{\rm opt}=0.85$. In Fig 5 the function $K_{\rm G-s}(\eta)|_{s=0.85}$ is plotted together with $K_{\rm G-s}(1/\eta)|_{s=0.85}.$
\begin{figure}[t]
  \centering
  \includegraphics[width=6cm]{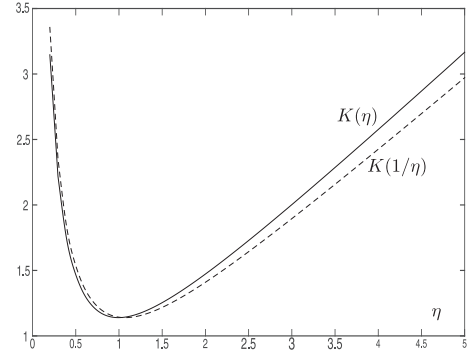}
  \caption{Generated by the Gauss-sinc wave function (\ref{Psi-s}) with $s=0.85$,  numerically calculated dependencies on $\eta$ (solid line) and on $1/\eta$ (dash line) of the Schmidt entanglement parameter $K_{G-s}$, the momentum representation.}\label{Fig5}
\end{figure}
As seen, the symmetry of the curves is not complete, but this is the best result attainable with a single fitting parameter $s$.

In Fig.6 we plot the same function $K_{\rm G-s}(\eta)|_{s=0.85}$  together with the Schmidt parameter $K_{2G}(\eta)$ (\ref{K-2G}) and  with two other entanglement parameters,  $R_{\rm diag}(\eta)$ and $R_{\rm tr}(\eta)$ found numerically with the same optimized wave function $\psi_s$ (\ref{Psi-s}) and  $s=0.85$.
\begin{figure}[h]
  \centering
  \includegraphics[width=8cm]{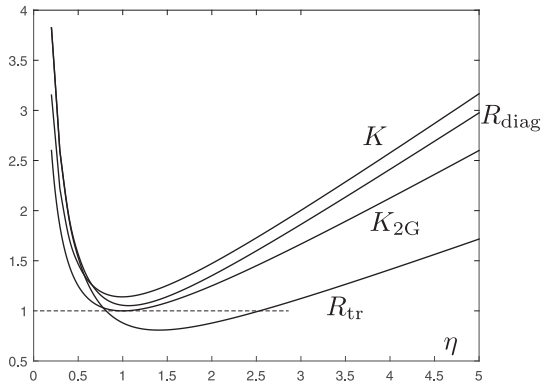}
  \caption{Transverse-entanglement parameters in the momentum representation vs the control parameter $\eta$ (\ref{eta}): The Schmidt parameter $K$, the diagonal and traditional width-ratio parameters $R_{\rm diag}$ and $R_{\rm tr}$ and, for comparison, the $2G$ Schmidt parameter $K_{\rm 2G}$ (\ref{K-2G}).}\label{Fig6}
\end{figure}
By comparing all these parameters with $K_{G-s}(\eta)$ we can draw the following conclusions: 1) The diagonal entanglement parameter $R_{\rm diag}(\eta)$ is systematically close to $K_{\rm G-s}(\eta)$and, in any case much closer than any other parameters. 2) The state characterized by the Gauss-sinc wave function (\ref{wf-k}), (\ref{Psi-s}) is always entangled, even at $\eta=1$ where $K_{\rm G-s}(\eta=1)>1$ and $R_{\rm diag}(\eta=1)>1)$, whereas $K_{\rm 2G}(\eta=1)=1$ (no entanglement),  3) the double-Gaussian modeling systematically underestimates the degree of transverse entanglement as $K_{\rm 2G}(\eta)$ pronouncedly smaller than $K_{\rm G-s}(\eta)$. 4) the parameter $R_{\rm tr}(\eta)$ differs significantly from $K_{\rm G-s}(\eta)$ and $R_{\rm diag}(\eta))$. Besides, in a rather wide variation range of the control parameter $\eta$, $R_{\rm tr}(\eta)$ appears to be smaller than one, $R_{\rm tr}(\eta)<1$, which undermines in this case its physical interpretations as the ratio of widths of the wider unconditional to the narrower conditional distributions.

\begin{figure}[h]
  \centering
  \includegraphics[width=6cm]{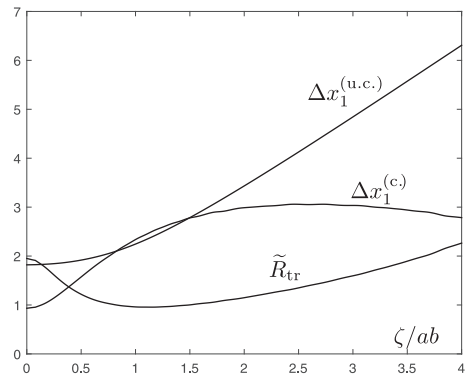}
  \caption{The widths of the conditional and unconditional distributions in the coordinate representation vs. the dimensionless propagation distance, and the parameter ${\widetilde R}_{\rm tr}(\zeta/ab)$}.\label{Fig7}
\end{figure}

To further analyze the properties of the entanglement parameters in the coordinate representation (without Gaussian modeling), we numerically performed the Fourier transform of the wave function $\psi_s(k_{1x},k_{2x})$ (\ref{Psi-s}) with $s=0.85$ and with added propagation factor $e^{-\zeta(k_{1x}^2+ k_{2x}^2)}$, and thus obtained the coordinate-dependent wave function ${\widetilde\Psi}(x_1, x_2; \zeta)$, unfortunately,  having no simple analytical representation. Nevertheless, directly with the obtained coordinate-dependent wave function we found numerically the conditional and unconditional single-particle distributions $|{\widetilde\Psi}(x_1, 0; \zeta)|^2$ and $\int dx_2|{\widetilde\Psi}(x_1,x_2; \zeta)|^2$, their widths $\Delta x_1^{\rm (c.)}(\zeta)$ and $\Delta x_1^{\rm (u.c.)}(\zeta)$, and the ratio of widths - the traditional entanglement parameter ${\widetilde R}_{\rm tr}=\Delta x_1^{\rm (u.c.)}/\Delta x_1^{\rm (c.)}$, which are shown in Fig. \ref{Fig7}.

The curves in Fig. \ref{Fig7} are similar to those of Fig. \ref{Fig3} and to the results of the work \cite{MVChe}, and they confirm that beyond the near zone ($\zeta\neq 0$) the traditional width-ratio parameter ${\widetilde R}_{\rm tr}(\zeta)$ fails to characterize the degree of transverse  entanglement because it strongly deviates from the Schmidt entanglement parameter $K(\zeta)={\rm const.}$.

\begin{figure}[h]
  \centering
  \includegraphics[width=6cm]{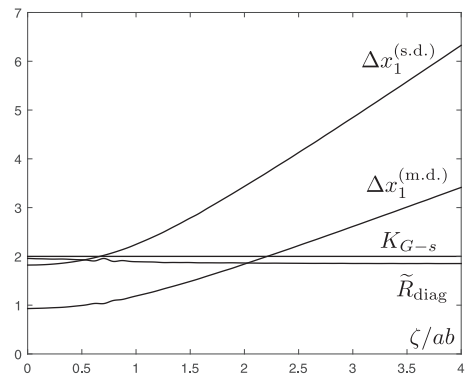}
  \caption{Widths of the conditional and unconditional distributions in the coordinate representation vs. the propagation distance, and the parameter ${\widetilde R}_{\rm diag}(\zeta/ab)$}.\label{Fig8}
\end{figure}

The second result which can be obtained from the calculated numerically  coordinate-dependent wave function ${\widetilde\Psi}(x_1, x_2; \zeta)$ concerns the diagonal distributions and diagonal entanglement parameter. The side-diagonal single-particle distribution and its width coincide with the discussed above unconditional distribution and width:  $\Delta x_1^{\rm (s.d.)}\equiv\Delta x_1^{\rm (u.c.)}$. As for the main-diagonal distribution, it is given by $\int dx_2{\widetilde\Psi}(x_1,x_2; \zeta){\widetilde\Psi}^*(x_1,-x_2; \zeta)$. The integral in this expression  also was calculated numerically, as well as the widths $\Delta x_1^{\rm (m.d.)}$, as well as the ratio of the widths ${\widetilde R}_{\rm diag}$, and the results are shown in Fig. \ref{Fig8}.

As seen, both widths $\Delta x_1^{\rm (s.d.)}(\zeta)$ and $\Delta x_1^{\rm (m.d.)}(\zeta)$ are monotonously growing functions, but their ratio ${\widetilde R}(\zeta)$ remains almost constant, and the value of ${\widetilde R}_{\rm diag}$ is very close to that of the Schmidt parameter $K_{G-s}$. This result confirms the conclusion that as with the Gaussian modeling or without any modeling the diagonal entanglement parameter is a good transverse entanglement quantifier alongside with the Schmidt entanglement parameter both in the near zone and beyond it.

\section{Possibilities of measurements}

For the state characterized by the wave function $\Psi(x_1, x_2;\zeta)$ its reduced density matrix is given by
\begin{gather}
\nonumber
 \rho_r(x_1,x_1^\prime;\zeta)=\int dx_2{\widetilde\Psi}(x_1, x_2;\zeta){\widetilde\Psi}^*(x_1^\prime, x_2;\zeta)=\\
 \nonumber
 =\int dx_2|{\widetilde\Psi}(x_1,x_2;\zeta)||{\widetilde \Psi}^*(x_1^\prime,x_2;\zeta)|\times\\
 \label{rho-r-gen}
 \times e^{i[\varphi(x_1,x_2;\zeta)-\varphi(x_1^\prime,x_2;\zeta)]},
\end{gather}
where $\varphi(x_1,x_2;\zeta)$ is the phase of $\Psi(x_1, x_2;\zeta)$.

Absolute values of the coordinate wave function, $|{\widetilde\Psi}(x_1,x_2;\zeta)|$, can be measured rather easily in a traditional way of splitting the beam of biphotons for two channels and measuring coincidence signals by two detector, one in the upper and one in the lower channels, with varying locations of detectors, which gives
\begin{equation}
 \label{counts}
 |{\widetilde\Psi}(x_1,x_2;\zeta)|=\sqrt{n(x_1,x_2;\zeta)},
\end{equation}
where $n(x_1,x_2;\zeta)$ is the relative number of detector counts during some given time and with given  locations  of detectors at $x_1$ and $x_2$. These measurements are sufficient for finding the side-diagonal single-particle distribution
\begin{equation}
 \label{single-dist via counts}
 \rho_r(x_1,x_1)=\frac{dW^{\rm(s.d.)}}{dx_1}=\sum_{x_2}n(x_1,x_2;\zeta),
\end{equation}
where the sum over $x_2$ imitates integration and $\sum_{x_1, x_2}n(x_1,x_2;\zeta)=1$.

As for the main-diagonal single-particle distribution determined by the reduced density matrix,in terms of measured absolute values of the wave function (\ref{counts}) it is given by
\begin{gather}
 \nonumber
 \rho_r(x_1,-x_1;\zeta)=\sum_{x_2}\sqrt{n(x_1,x_2;\zeta)n(-x_1,x_2;\zeta)}\times\\
 \label{reconstr}
 \times\exp\left\{i\left[\varphi(x_1,x_2;\zeta)-
 \varphi(-x_1,x_2;\zeta)\right]\right\}.
\end{gather}
Clearly, in this case measurement of phases of the wave function is needed and is unavoidable.

Such measurements hardly can be done in any simple way. In principle, various aspects of similar problems were discussed  in a number of works \cite{LR, Mukam, RWB, Walborn, Zou, Lun, Landon, Villoresi}. Not pretending for giving here an overview of these works or adapting any of their methods to our goals, we describe below a scheme of measurements which seems to be appropriate for finding phases of the wave function under consideration. The main idea of the scheme we suggest is similar to that of the work by Z.Y. Ou \cite{Ou}, which was aimed for investigations in the field of four-photon interference and provided selection of entangled four-photon polarization states from the manifold of independently born pairs of photons.

\begin{figure}[h]
  \centering
  \includegraphics[width=8cm]{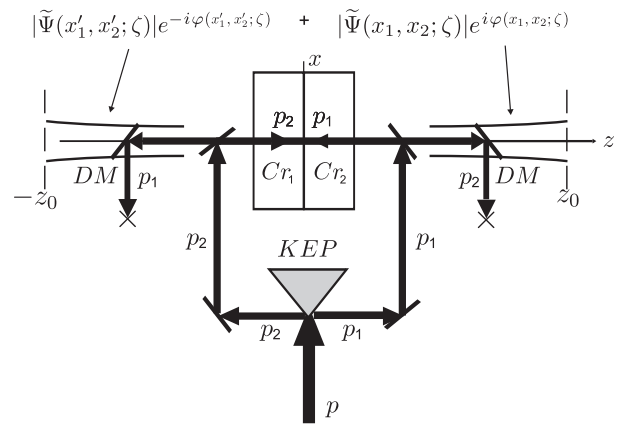}
  \caption{A scheme for measuring the phase  $\varphi(x_1,x_2;\zeta)$ of the wave function ${\widetilde \Psi}(x_1,x_2;\zeta)$ (part 1); $KEP$ is the knife-edge prism splitting the pump ($p$) for  two equal counter propagating parts, $Cr$ and $DM$ denote two identical nonlinear crystals and dichroic mirrors; in principle, the $KEP$ can be replaced by a beamsplitter and a couple of additional mirrors. \label{Fig9}}
\end{figure}

The scheme we suggest is shown in Fig. \ref{Fig9}.

\begin{figure}[h]
  \centering
  \includegraphics[width=8cm]{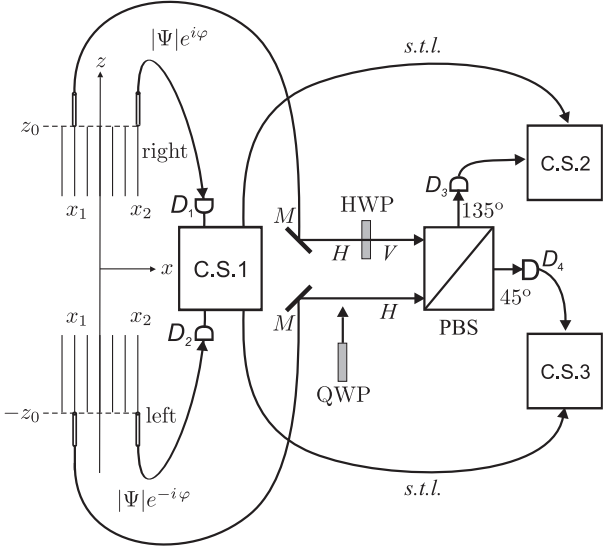}
  \caption{A scheme for extracting photons and measuring the phase $\varphi(x_1,x_2;\zeta)\equiv\varphi$ (part 2); C.S.1, C.S.2 and C.S.3 are coincidence schemes, $D_1$-$D_4$ are detectors, PBS is the polarization beamsplitter turned for $45^{\rm o}$ around the $x$-axis, HWP and QWP are the half-wave and quarter-wave phase plates, with the latter to be installed into the lower channel for the 2nd series of measurements; $s.t.l.$ are the signal transmission lines from C.S.1 to SC.S.2 and C.S.3}\label{Fig10}.
\end{figure}

In this scheme the pump ($p$) is assumed to be split for two equal counter-propagating parts with equal intensities, $p_1$ and $p_2$. Each part of the split pump is directed to the pair of crystals ${Cr_1}$ and ${Cr_2}$ from opposite sides, as shown in Fig. \ref{Fig9}. The crystals ${Cr_1}$ and ${Cr_2}$ are assumed to be identical but slightly differently oriented. The crystal ${Cr_1}$ is assumed to provide  the phase matching conditions only for the pump $p_1$ and SPDS photons propagating into the direction of negative $z$ but not for the pump $p_2$ propagating into direction of positive $z$. Oppositely, the pump $p_2$ propagates through the crystal $Cr_1$ in the direction to positive $z$ without any effects or changes and begins stimulating production of SPDC photons only when it comes to the crystal $Cr_2$. The total wave function of biphoton pairs produced in two crystals if given by the sum of contributions from the crystal $Cr_1$ and $Cr_2$. In principle, pairs produced in two crystals are not necessarily born simultaneously. But for measuring phases, the scheme of measurements has to select only pairs simultaneously born in both crystals. Such pairs are characterized by the wave functions with equal absolute values and different signs of phases as shown on the top of Fig. \ref{Fig9}. Such scheme of measurements is shown in Fig. \ref{Fig10}, which is a direct continuation of Fig.\ref{Fig9}, though turned for $90^{\rm o}$ for convenience

Photons are assumed to be extracted from both shoulders by thin pieces of fibers parallel to the $z-$axis at the same distances $x_1$ and $x_2$ from the central propagation axis $z$ in the left and right shoulders of the scheme of Fig. \ref{Fig9}. The total wave function of extracted photons is given by the sum of contributions from the left and right crystals
\begin{equation}
 \label{tot wf}
 \Psi_{\rm extr}=|\Psi(x_1,x_2;\zeta)|\left[e^{-\varphi(x_1,x_2)}+e^{i\varphi(x_1,x_2)}\right].
\end{equation}
The extracted photons propagate in fibers along short ($x_2$-photons) and long ($x_1$-photons) trajectories. The short-route photons move to the detectors $D_1$ and $D_2$ and then to the coincidence scheme $C.S.1$ which registers only $x_2$-photons born simultaneously in the left and right shoulders. Immediately after registration in the coincidence scheme $C.S.1$, signals are sent to the coincidence schemes $C.S.2$ and $C.S.3$ destined for registration of pairs of long-route photons. These photons are assumed to be analyzed with the help of the polarization beamsplitter turned for 45$^{\rm o}$ around the photon propagation $x$-axis. But before this, horizontal polarization of photons in one of two channels has to be changed from  horizontal to vertical with the help of the half-wave phase plate (HWP) with the optical axis directed along the bisector between the horizontal and vertical directions. After this, the quantum state of two long-route photons can be characterized by the state vector
\begin{gather}
 \nonumber
 \ket{\psi_{x_1}}=\frac{1}{\sqrt{2}}\left (a_H^\dag e^{-i\varphi}+a_V^\dag e^{i\varphi}\right ) \ket{0}=\\
 \label{two-phot}
 (-\sin\varphi a_{45^{\rm o}}^\dag+\cos\varphi a_{135^{\rm o}}^\dag)\ket{0},
\end{gather}
where $a_H^\dag$ and $a_V^\dag$ are creation operators of photons with horizontal and vertical polarizations, whereas $a_{45^{\rm o}}^\dag$ and $a_{135^{\rm o}}^\dag$ are the creation operators of photons with polarizations along the directions at angles $45^{\rm o}$ and $135^{\rm o}$ with respect to the horizontal direction. The coefficients in front of the operators $a_{45^{\rm o}}^\dag$ and $a_{135^{\rm o}}^\dag$ are probability amplitudes of finding photons with these polarizations at the exit from PBS in the Fig.\ref{Fig10}. The corresponding probabilities are
\begin{equation}
 \label{prob}
 w_{45^{\rm o}}=\cos^2\varphi,\quad w_{135^{\rm o}}=\sin^2\varphi,
\end{equation}
and, hence,
\begin{equation}
 \label{cos}
 \cos (2\varphi)= w_{45^{\rm o}}- w_{135^{\rm o}}.
\end{equation}

Note that, as usual, probabilities $w_{45^{\rm o}}$ and $w_{135^{\rm o}}$ are determined by relative numbers of photons with polarizations $45^{\rm o}$ and $135^{\rm o}$ registered in the coincidence schemes C.S.2 and C.S.3. during some given time :
\begin{equation}
 \label{def of prob}
w_{45^{\rm o}}=\frac{N_{45^{\rm o}}}{N_{45^{\rm o}}+N_{135^{\rm o}}},\quad
w_{135^{\rm o}}=\frac{N_{135^{\rm o}}}{N_{45^{\rm o}}+N_{135^{\rm o}}}.
\end{equation}

Note also, that that after the detectors ${\rm D}_3$ and ${\rm D}_4$  accepting, correspondingly, the  $45^{\rm o}$- and $135^{\rm o}$-polarized photons, signals from these detectors are assumed to be sent to two different coincidence schemes, C.S.2 and and C.S.3. Other signals coming to these coincidence schemes are those coming via two transmission lines from the first coincidence scheme C.S.1, which selects simultaneous registration of two  short-route $x_1-$photons. The lengths of the transmission lines are assumed to be carefully adjusted to provide equal arrival times of both signals coming to the coincidence schemes C.S.2 and to C.S.3. The conditions of coincidences in these schemes select only events consisting in simultaneous birth of two pairs of photons with transverse coordinates $x_1$ and $x_2$, one pair in the left- and one pair in the right-hand crystals of the scheme in Fig. \ref{Fig9}.

Thus, Eqs. (\ref{cos}) and (\ref{def of prob}) show how $\cos2\varphi$ is related to experimentally measurable probabilities or numbers of events to be registered in the coincidence schemes C.S.2 and and C.S.3. But this is still not enough to unambiguously determine the phase $\varphi$, which requires at least an independent measurement of $\sin2\varphi$ in addition to $\cos2\varphi $. This problem can be solved quite simply by means of repeated measurements of the same type as above, but with a slight change in the  photon's polarization in the lower channel in front of the PBS in Fig. \ref{Fig10}. The required change can be produced by means of installation in the lower channel of the quarter-wave phase plate with the vertically oriented optical axis,which adds the imaginary unit in front of $a_V^\dag$ in  the quantum state of photons in front of PBS, which takes the form
\begin{gather}
 \nonumber
 \ket{{\widetilde\psi}_{x_1}}=\frac{1}{\sqrt{2}}  \left(i a_H^\dag e^{-i\varphi}+a_V^\dag e^{i\varphi}\right)\ket{0}=\\
 \label{QWP-modif}
 \left[\frac{i\,e^{-i\varphi}+e^{i\varphi}}{2}\,a_{45^{\rm o}}^\dag+
 \frac{-i\,e^{-i\varphi}+e^{i\varphi}}{2}\,a_{135^{\rm o}}^\dag\right]\ket{0}.
\end{gather}
The probabilities of registering photons with $45^{\rm o}$ and $135^{\rm o}$ are given by
\begin{equation}
 \label{45-tilde}
 {\widetilde w}_{45^{\rm o}}=\frac{\left|i\,e^{i\varphi}+e^{i\varphi}\right|^2}{4}=\frac{1+\sin 2\varphi}{2},
\end{equation}
and
\begin{equation}
 \label{135-tilde}
  {\widetilde w}_{135^{\rm o}}=\frac{\left|-i\,e^{i\varphi}+e^{i\varphi}\right|^2}{4}
 =\frac{1-\sin 2\varphi}{2},
\end{equation}
and the required expression for $\sin2\varphi$:
\begin{equation}
\label{sin}
 \sin2\varphi={\widetilde w}_{45^{\rm o}}-{\widetilde w}_{135^{\rm o}}
\end{equation}

\begin{figure}[t]
  \centering
  \includegraphics[width=8cm]{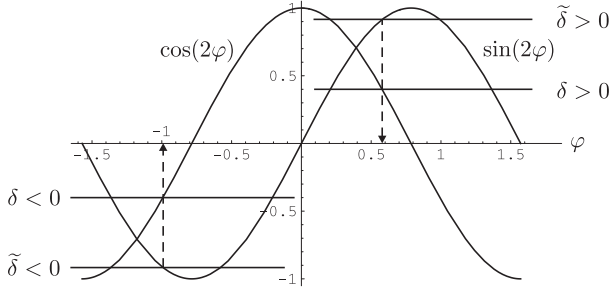}
  \caption{Indicated by dashed lines with arrows, solutions of the couple of equations (\ref{cos}) and (\ref{sin}) in the cases $\delta>0,\,{\widetilde \delta}>0$ and $\delta<0,\,{\widetilde \delta}<0$, where $\delta=w_{45^{\rm o}}-w_{135^{\rm o}}$ and ${\widetilde \delta}={\widetilde w}_{45^{\rm o}}-{\widetilde w}_{135^{\rm o}}$.} \label{Fig11}
\end{figure}

By assuming now that $|\varphi|\leq\pi/2$, we can find the following explicit expression for phase
$\varphi$ in terms of probabilities $w_{45^{\rm o}}$, $w_{135^{\rm o}}$, ${\widetilde w}_{45^{\rm o}}$ and ${\widetilde w}_{135^{\rm o}}$:
\begin{gather}
 \nonumber
  2\varphi=\arccos(w_{45^{\rm o}}-w_{135^{\rm o}}){\rm Sign}(\sin[2\varphi])=\\
 \label{phase}
 \arccos(w_{45^{\rm o}}-w_{135^{\rm o}}){\rm Sign}({\widetilde w}_{45^{\rm o}}-{\widetilde w}_{135^{\rm o}}),
\end{gather}
which is illustrated by the picture of Fig.\ref{Fig11}.

 In principle, the described measurements of the phase $\varphi(x_1,x_2;\zeta)$, repeated many times with various values of $x_1$ and $x_2$ and recorded to the computer memory, can be used for reconstruction of of the main-diagonal single-particle distribution (\ref{reconstr}). Together with the much easier found side-diagonal distribution (\ref{single-dist via counts}), this provides a way for direct experimental measurement of the diagonal entanglement parameter ${\widetilde R}_{\rm diag}$ (\ref{R diag}).

\section{Conclusion}

To conclude, let us summarize briefly the main derived results of the work.

1. It's proved that the reduced density matrix of a transverse biphoton state beyond the near zone  is real on and only on diagonals in the plane of transverse photon's coordinates, which makes the diagonal directions in this plane very special and peculiar. The given proof is rather general and is based in fact only on properties (\ref{identities}) of the near-zone momentum-representation wave function $\psi(k_{1x}, k_{2x})$ (\ref{wf-k}). These properties and the proof itself remain valid for many other forms of the wave function $\psi$. E.g. for the cases of the Gaussian pump profile replaced by the super-Gaussian one, or by the Lorentian form, etc., and the same with the sinc-function.

2. Because of the missing imaginary parts in the main- and side-diagonal elements of the reduced density matrix, they are interpreted as two orthogonal and complementary single-particle distributions $\rho_r(x_1,x_1)$ and $\rho_r(x_1,-x_1)$, widths of which can be used for the definition of a new, fully diagonal entanglement parameter ${\widetilde R}_{\rm diag}$. Fruitfulness of this idea is confirmed by direct numerical calculations for the case of the true biphoton wave function $\psi(k_{1x}, k_{2x})$ (\ref{wf-k}) without any modeling. The results of calculation presented in Fig. \ref{Fig8} show that, indeed, the parameter ${\widetilde R}_{\rm diag}$ is very close to the Schmidt parameter $K$ and almost is not affected by the diffraction spreading of the biphoton beam beyond the near zone.

We believe that these results provide rather important and interesting new knowledge about features of the reduced density matrix of the transverse-variable biphoton state beyond the near zone and about parameters characterizing the degree of its entanglement.

\bibliography{Trans-2A}

\end{document}